\documentclass[pdflatex,sn-basic]{sn-jnl} % "referee" for double spacing

\jyear{2025}%

\theoremstyle{thmstyleone}

\theoremstyle{thmstyletwo}

\theoremstyle{thmstylethree}

\usepackage{amssymb}
\usepackage{pifont}

\usepackage{amsmath}
\usepackage{lineno}
\usepackage{accsupp}
\usepackage{fancyhdr}
\usepackage{lastpage}
\usepackage{adjustbox}
\usepackage{tabularx}
\usepackage{makecell}
\usepackage{multirow}

\begin{document}

\title[Article Title]{Rapid updating of multivariate resource models based on new information using EnKF-MDA and multi-Gaussian transformation}

\author*[1,2]{\fnm{Sultan} \sur{Abulkhair}}\email{sultan.abulkhair@adelaide.edu.au}

\author[1,2]{\fnm{Peter} \sur{Dowd}}\nomail

\author[2]{\fnm{Chaoshui} \sur{Xu}}\nomail

\author[3]{\fnm{Penny} \sur{Stewart}}\nomail

\affil[1]{\orgdiv{ARC Training Centre for Integrated Operations for Complex Resources}, \orgname{The University of Adelaide}, \orgaddress{\city{Adelaide, SA 5005}, \country{Australia}}}

\affil[2]{\orgdiv{School of Chemical Engineering}, \orgname{The University of Adelaide}, \orgaddress{\city{Adelaide, SA 5005}, \country{Australia}}}

\affil[3]{\orgname{Petra Data Science}, \orgaddress{\city{Brisbane Qld 4000}, \country{Australia}}}

\abstract{Rapid resource model updating with real-time data is important for making timely decisions in resource management and mining operations. This requires optimal merging of models and observations, which can be achieved through data assimilation, and the ensemble Kalman filter (EnKF) has become a popular method for this task. However, the modelled resources in mining usually consist of multiple variables of interest with multivariate relationships of varying complexity. EnKF is not a multivariate approach, and even for univariate cases, there may be slight deviations between its outcomes and observations. This study presents a methodology for rapidly updating multivariate resource models using the EnKF with multiple data assimilations (EnKF-MDA) combined with rotation based iterative Gaussianisation (RBIG). EnKF-MDA improves the updating by assimilating the same data multiple times with an inflated measurement error, while RBIG quickly transforms the data into multi-Gaussian factors. The application of the proposed algorithm is validated by a real case study with nine cross-correlated variables. The combination of EnKF-MDA and RBIG successfully improves the accuracy of resource model updates, minimises uncertainty, and preserves the multivariate relationships.}


\maketitle

\section{Introduction}\label{sec1}

Resource modelling is a crucial component of the mining value chain, as it outlines the quantity, quality, and location of mineral resources within a specified area. However, resource models are typically based on limited exploration data collected over a large area, often failing to accurately represent reality. As these models are the foundation for future planning, predictions, and optimisations, their accuracy directly impacts production outcomes. To address potential discrepancies, it is essential to evaluate risks by quantifying the uncertainty associated with the orebody using advanced geostatistical simulation algorithms \citep{bib1}. Additionally, collecting more direct measurements during operations can enhance the accuracy and precision of these models.

Real-time production data from various sensors can quickly update resource knowledge and inform short-term mine planning decisions. Incorporating these observations into geostatistical models can enhance the accuracy and precision of forecasts. \cite{bib2} suggest that integrating sensor data could reduce uncertainty and deviations from production targets, potentially resulting in an average economic benefit of \$5 million per year for the studied deposit.

\begin{figure}[ht]%
\centering
\includegraphics[width=0.8\textwidth]{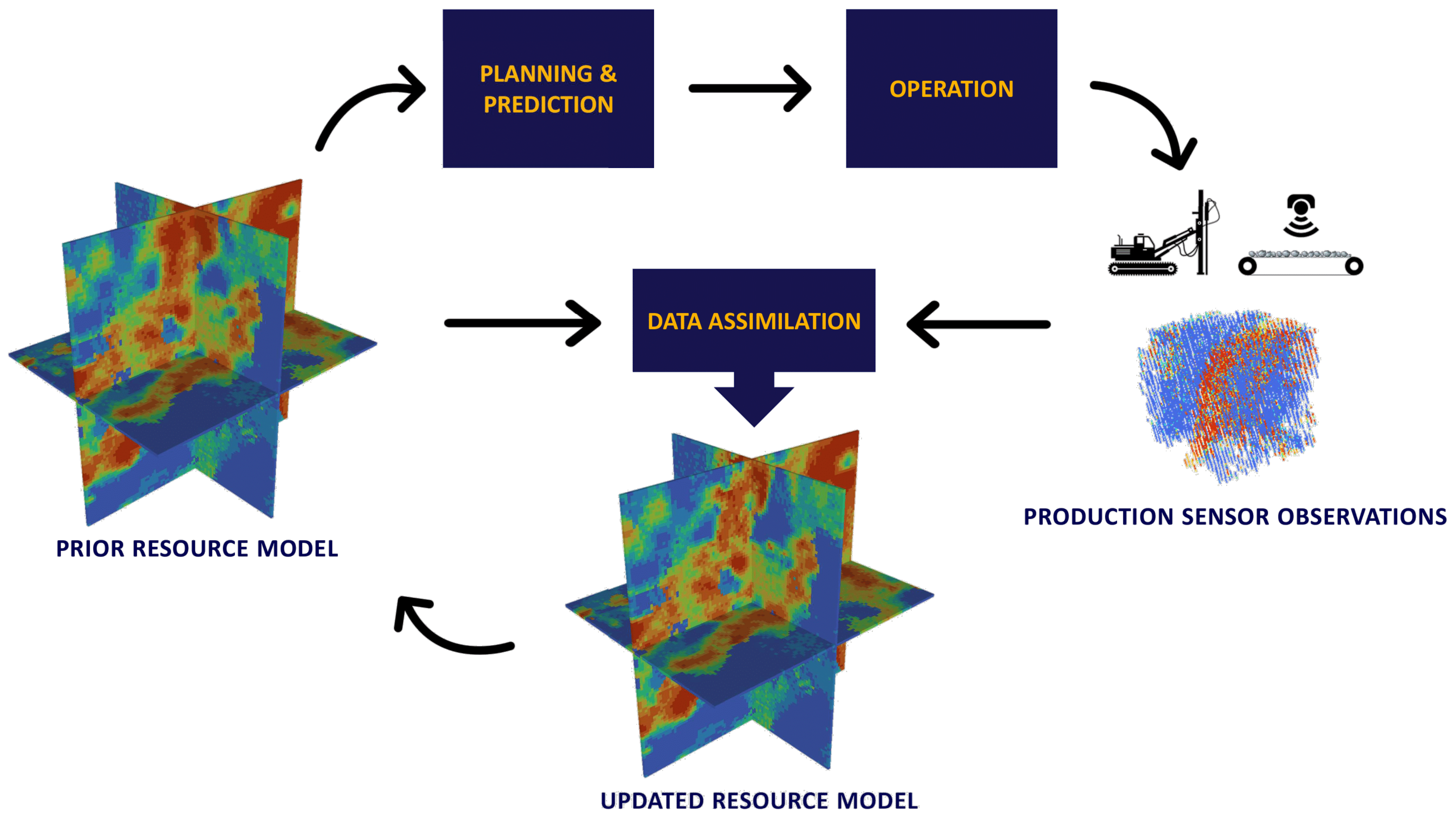}
\caption{Real-time mining concept inspired by \cite{bib7}.}\label{fig1}
\end{figure}

To effectively integrate sensor observations, tools such as the Kalman filter (KF) \citep{bib3} and the ensemble Kalman filter (EnKF) \citep{bib4} are recommended for the rapid updating of resource and grade control models \citep{bib5,bib6}. In addition, \cite{bib7} introduced a real-time mining concept that transforms discontinuous process monitoring into a near-continuous framework through data assimilation (see Figure~\ref{fig1}). Data assimilation applications extend beyond mineral grade estimates and can also be used to update coal quality parameters \citep{bib8}, geometallurgical models \citep{bib9}, and compositional data \citep{bib10}. However, current data assimilation methods for quickly updating resource models still have limitations. For instance, the EnKF updates one variable at a time and does not consider multivariate relationships. Furthermore, single data assimilation is often insufficient to fully utilise the information obtained from observations.

One practical approach for performing multivariate rapid updating is to transform co-regionalised variables into independent factors before data assimilation. This transformation helps preserve multivariate relationships by back-transforming realisations after updates. Techniques such as the minimum-maximum autocorrelation factors (MAF) \citep{bib11} and flow transformation (FA) \citep{bib12} have been used alongside the EnKF to decorrelate multivariate data before rapid updating \citep{bib13,bib10}. However, MAF is not well-suited for handling complex multivariate relationships, and FA tends to be too slow for adequate rapid updating. In a comparison of various multivariate transformations, \cite{bib14} found that the projection pursuit multivariate transform (PPMT) \citep{bib15} and rotation-based iterative Gaussianisation (RBIG) \citep{bib16} are considerably faster and more appropriate for rapid updating than FA. However, it should be noted that only FA can successfully minimise artefacts in the presence of extreme values, making it a more optimal multi-Gaussian transform, provided that runtime is not important \citep{bib17}.

RBIG is similar in its methodology to PPMT, but instead of iteratively searching for interesting projections, it applies orthonormal rotations. As it needs fewer iterations to achieve convergence and each iteration is faster, RBIG is, on average, 90\% faster than PPMT when applied to five variables with the number of samples between 10,000 and 50,000 \citep{bib17}. Similar to PPMT and FA, RBIG can be paired with other transforms, such as MAF \citep{bib18,bib14}, log-ratio transforms \citep{bib12,bib10}, and non-logarithmic ratio transforms \citep{bib19,bib20}.

In addition to the EnKF, several other methods have been employed for the rapid updating of resource models, including conditional simulation of successive residuals \citep{bib21}, direct sequential simulation using point distributions \citep{bib22}, a variation of the KF for downscaling resource models \citep{bib23}, and deep reinforcement learning \citep{bib24}. Most of these methods perform single data assimilation, which can lead to some deviations between predicted and observed values, especially when observations are highly uncertain. To improve data assimilation in highly non-linear situations, iterative forms of EnKF have been widely explored in petroleum engineering and hydrogeology. For instance, ensemble randomised maximum likelihood, also known as iterative EnKF \citep{bib25,bib26}, yields better-matching results than standard EnKF, although it requires more computational resources. Alternatively, EnKF with multiple data assimilations (EnKF-MDA) \citep{bib27,bib28} performs multiple data assimilations with an inflated measurement error to significantly outperform single data assimilation methods. While both iterative EnKF and EnKF-MDA are more computationally intensive than standard EnKF, the EnKF-MDA approach typically requires fewer iterations. The effectiveness of EnKF-MDA has been demonstrated in hydrogeology \citep{bib29}, petroleum engineering \citep{bib27,bib28} and geophysics \citep{bib30}. However, to our knowledge, EnKF-MDA has yet to be applied to the rapid resource model updating problem in mining.

Validating the performance of data assimilation algorithms through real case studies is crucial, especially in the mining industry, where data often show significant variability in terms of spacing between data points, measurement volumes, and associated uncertainties. Several studies of rapid resource model updating have demonstrated the effectiveness of various proposed approaches when applied to real data \citep{bib32,bib9,bib13,bib31,bib10}. However, some of these studies sample observations from a ground truth model generated from real data \citep{bib31,bib10}. This makes it difficult to account for preferential sampling and incoming new observations from previously under-sampled locations. Both of these problems complicate multivariate transformations since using the same transformation function for model realisations and new observations may become unreliable.

In this paper, we apply a combination of EnKF-MDA and RBIG to a real case study from an iron ore mine in Western Australia. Because of its iterative nature, EnKF-MDA achieves a better match between observations and model-based predictions than EnKF. Additionally, RBIG enables EnKF-MDA to account for multivariate relationships between cross-correlated variables. Because the real data were fused to track historical data to the resource model locations, it makes it easier to validate the proposed approach.

The following section offers a detailed methodology for the proposed multivariate rapid updating algorithm. Next, real fused data is used to sequentially update the resource model over 25 time periods, the results of which are thoroughly analysed. The paper then concludes with a discussion of key results, limitations of the proposed algorithm, and future research directions.

\section{Methods}\label{sec2}

\subsection{Rapid updating of multivariate resource models}\label{subsec2.1}

The proposed updating approach uses EnKF-MDA for data assimilation, paired with RBIG for decorrelation and Gaussianisation of cross-correlated variables. The steps in the proposed algorithm are as follows:

\begin{enumerate}[1.]

\item Select a neighbourhood around the observations and extract block model realisations located within that neighbourhood. In this paper, the neighbourhood is determined by the pre-defined number of blocks from each observation.

\item Transform neighbourhood realisations and new observations into multi-Gaussian factors using RBIG
\begin{equation}
\left({Z_{e}^{n,t}}^{\text{RBIG}},{O_{e}^{t}}^{\text{RBIG}}\right)=\Phi_{\text{RBIG}}^{t}\left(Z_{e}^{n,t},O_{e}^{t}\right),\label{eq1}
\end{equation}
where $Z_{e}^{n,t}$ are $n$ neighbourhood realisations, $O_{e}^{t}$ are observations with a number of variables $e$ at a time $t$, and $\left({Z_{e}^{n,t}}^{\text{RBIG}},{O_{e}^{t}}^{\text{RBIG}}\right)$ are multi-Gaussian realisations and observations.

\item Apply EnKF-MDA to get updated multi-Gaussian realisations ${Z_{e}^{n,t+1}}^{\text{RBIG}}$ based on prior realisations and observations
\begin{equation}
{Z_{e}^{n,t+1}}^\text{RBIG}=\Phi_{\text{EnKF-MDA}}^{t}\left({Z_{e}^{n,t}}^{\text{RBIG}},{O_{e}^{t}}^{\text{RBIG}}\right).\label{eq2}
\end{equation}

\item Back-transform updated realisations into the original state
\begin{equation}
Z_{e}^{n,t+1}={\Phi_{\text{RBIG}}^{t}}^{-1}\left({Z_{e}^{n,t+1}}^\text{RBIG}\right),\label{eq3}
\end{equation}
where ${\Phi_{\text{RBIG}}^{t}}^{-1}$ is an inverse RBIG transformation and $Z_{e}^{n,t+1}$ are back-transformed updated neighbourhood realisations.

\item Insert the updated neighbourhood realisations back into the block model. 

\end{enumerate}

\subsection{Rotation based iterative Gaussianisation}\label{subsec2.2}

RBIG is an iterative multi-Gaussian transform based on marginal Gaussianisations followed by orthonormal rotations \citep{bib16}. RBIG is chosen as the optimal multi-Gaussian transform for rapid updating because it is significantly faster than other methods while maintaining comparable performance across various metrics \citep{bib14}. Principal component analysis (PCA) and independent component analysis (ICA) can be used for orthonormal rotations. Nevertheless, RBIG achieves successful convergence regardless of the choice of orthonormal rotations and although ICA requires fewer iterations, each of its iterations is more computationally complex than those in PCA \citep{bib16}. ICA also generates some significant artefacts after back-transformation, whereas RBIG with PCA produces results similar to PPMT \citep{bib14}.

In this paper, PCA is chosen for orthonormal rotations. A single iteration of RBIG is defined as

\begin{equation}
Y_{e}^{i+1}=R^{i}\Psi^{i}\left(X_{e}^{i}\right),\label{eq4}
\end{equation}
where $\Psi^{i}\left(X_{e}^{i}\right)$ is a marginal Gaussianisation based on histogram equalisation of the multivariate data at iteration $i$ and $R^{i}$ is a PCA rotation matrix.

Marginal Gaussianisation functions and rotation matrices at each iteration are stored as $\Phi_{\text{RBIG}}^{t}$. The back-transformation ${\Phi_{\text{RBIG}}^{t}}^{-1}$ is performed by following the stored iterations in reverse order.

\subsection{Ensemble Kalman filter with multiple data assimilations}\label{subsec2.3}

EnKF-MDA is an iterative version of EnKF as it performs multiple data assimilations on the same data with an inflated measurement error \citep{bib27}. It provides a significantly better match between predictions and observations compared to single data assimilation methods while not being overly computationally intensive \citep{bib28}. The steps of the EnKF-MDA implemented for rapid resource model updating are the following:

\begin{enumerate}[1.]

\item Compute model-based predictions $H_{e}^{n,t}$ at observation locations.

\item Add random noise to observations according to the measurement error
\begin{equation}
O_{e}^{t}=O_{e}^{t}+\varepsilon.\label{eq5}
\end{equation}

\item Compute the Gaspari-Cohn correlation filter for covariance localisation
\begin{equation}
\alpha(r)=
\begin{cases}
-\frac{1}{4}r^{5}+\frac{1}{2}r^{4}+\frac{5}{8}r^{3}-\frac{5}{2}r^{2}+1 & 0\le r<1\\
\frac{1}{12}r^{5}-\frac{1}{2}r^{4}+\frac{5}{8}r^{3}+\frac{5}{3}r^{2}-5r+4-\frac{2}{3}r^{-1} & 1<r\le2\\
0 & \hphantom{.1<}r>2
\end{cases},\label{eq6}
\end{equation}
where $r$ is a normalised distance between two locations defined by $\frac{d}{L}$, $d$ is a distance and $L$ is a predefined localisation radius.

\item Compute the Kalman gain
\begin{equation}
K_{e}^{t}=\alpha(r)C_{YD}^{e,t}\left(\alpha(r)C_{DD}^{e,t}+C_{D}^{e,t}\right),\label{eq7}
\end{equation}
where $C_{YD}^{e,t}$ is the experimental covariance between realisations and model-based predictions, $C_{DD}^{e,t}$ is the experimental covariance of model-based predictions and $C_{D}^{e,t}$ is the experimental covariance of observations.

\item Update prior realisations
\begin{equation}
{Z_{e}^{n,t+1}}^\text{RBIG}={Z_{e}^{n,t}}^\text{RBIG}+K_{e}^{t}\left({O_{e}^{t}}^{\text{RBIG}}-H_{e}^{n,t}\right).\label{eq8}
\end{equation}

\item Repeat steps 1-5 for a predefined number of data assimilations and ensure that random noise added to observations differs from previous data assimilations. 

\end{enumerate}

\section{Results}\label{sec3}

\subsection{Overview of a case study}\label{subsec3.1}

In this case study, the proposed approach is applied to real anonymised mining data. The dataset was fused to track the historical data points back to their respective resource blocks. Readers are referred to the paper \citep{bib33} for more details about the data fusion and orebody learning involved in creating this dataset. Table~\ref{tab1} shows the statistics of the fused dataset, including the mean, standard deviation, skewness and kurtosis. The data are from an iron ore deposit in Western Australia and include the nine cross-correlated assay variables Fe, S, SiO$_2$, CaO, MgO, P, Al$_2$O$_3$, TiO$_2$ and K$_2$O. Note that the assay variables are not listed in the same order as the variable names in the table. Due to confidentiality reasons, we cannot disclose the name of the deposit, the variable names, or the coordinates.

\begin{table}[h]
\begin{center}
\caption{Statistical parameters of the original data.}\label{tab1}%
\begin{adjustbox}{width=1\textwidth}
\begin{tabular}{lccccccccc}
\toprule
Parameter & Assay 1 & Assay 2 & Assay 3 & Assay 4 & Assay 5 & Assay 6 & Assay 7 & Assay 8 & Assay 9\\
\midrule
Mean & 1.93 & 58.39 & 0.009 & 6.32 & 0.055 & 0.049 & 0.029 & 7.88 & 0.072\\
Standard deviation & 1.10 & 6.03 & 0.014 & 1.42 & 0.028 & 0.012 & 0.020 & 8.66 & 0.069\\
Skewness & 2.67 & -1.83 & 8.63 & 0.76 & 2.29 & 1.90 & 1.59 & 2.39 & 2.69\\
Kurtosis & 13.27 & 3.48 & 111.66 & 0.66 & 9.69 & 5.18 & 3.35 & 5.56 & 10.29\\
\botrule
\end{tabular}
\end{adjustbox}
\end{center}
\end{table}

\begin{figure}[ht]%
\centering
\includegraphics[width=\textwidth]{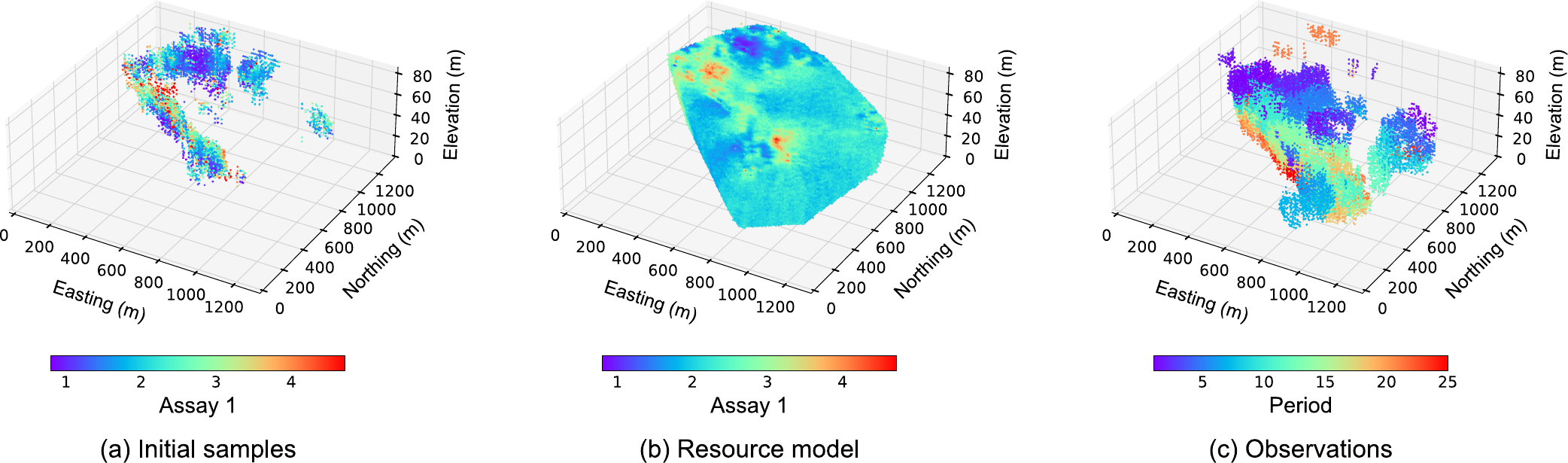}
\caption{3D view of initial samples of Assay 1, a prior resource model with average simulated values of Assay 1 and observations divided into 25 periods.}\label{fig2}
\end{figure}

The original data points are categorised into 31 periods based on the months in which they were collected. A total of 5,327 data points from the first six months were used to simulate the prior resource model. Turning bands simulation was used to model the multi-Gaussian factors of the original assay variables (the variogram models are shown in Table~\ref{tab2}). The remaining 25 periods serve as observations to sequentially update the resource model. Figure~\ref{fig2} provides a 3D view of the study area, specifically highlighting the initial six months of samples, the prior resource model, and the observations that are used to update the model. For simplicity, the measurement error is assumed to be a constant value of 10\%.

\begin{table}[h]
\begin{center}
\caption{Omni-directional direct variogram models of the multi-Gaussian factors.}\label{tab2}%
\begin{adjustbox}{width=0.7\textwidth}
\begin{tabular}{lc}
\toprule
Variable & Variogram model\\
\midrule
Assay 1 & 0.28Nug + 0.33Sph(37m) + 0.39Sph(358m)\\
Assay 2 & 0.30Nug + 0.44Sph(39m) + 0.26Sph(216m)\\
Assay 3 & 0.29Nug + 0.55Sph(47m) + 0.16Sph(494m)\\
Assay 4 & 0.38Nug + 0.34Sph(30m) + 0.28Sph(186m)\\
Assay 5 & 0.34Nug + 0.39Sph(30m) + 0.08Sph(211m) + 0.19Sph(213m)\\
Assay 6 & 0.30Nug + 0.24Sph(28m) + 0.22Sph(76m) + 0.24Sph(279m)\\
Assay 7 & 0.47Nug + 0.36Sph(55m) + 0.17Sph(318m)\\
Assay 8 & 0.35Nug + 0.38Sph(29m) + 0.27Sph(111m)\\
Assay 9 & 0.32Nug + 0.68Sph(56m)\\
\botrule
\end{tabular}
\end{adjustbox}
\end{center}
\end{table}

\begin{figure}[ht]%
\centering
\includegraphics[width=0.65\textwidth]{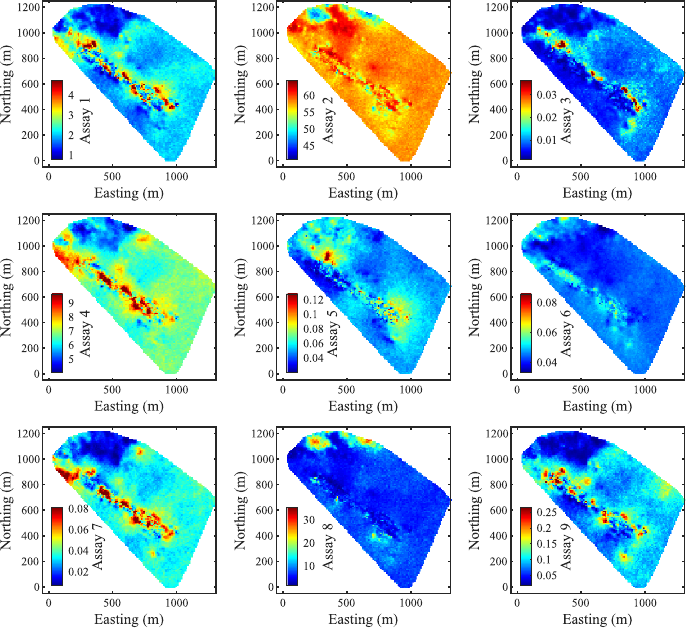}
\caption{2D view of prior resource models at 44 m elevation.}\label{fig3}
\end{figure}

The resource model consists of 100 realisations with blocks of size 10 m × 10 m × 4 m. Figure~\ref{fig3} provides a 2D view of the e-type models for each assay variable at an elevation of 44 m. As the observations were tracked back to their corresponding blocks, the spacing between the data points aligns with the dimensions of the blocks.

\subsection{An illustrative example of rapid updating for period 1}\label{subsec3.2}

As outlined in the methodology of the proposed updating algorithm, observations are not used to update the entire block model. For one reason, updating blocks that are far from the observations would unnecessarily reduce the uncertainty of those blocks. More importantly, calculating covariance matrices for the entire model is both time-consuming and inefficient in terms of memory usage. Therefore, the first step of the updating algorithm is to select the neighbourhood around the observations. In this study, the neighbourhood is defined as being within three blocks away from observation points. For example, for period 1, with 868 observations, the neighbourhood consists of 15,561 blocks.

The next step involves transforming neighbourhood realisations and observations into multi-Gaussian factors. If the prior model was created using RBIG, one possible strategy would be to use stored RBIG functions and matrices from the prior model. However, this could make the proposed algorithm inflexible and overly dependent on the original exploration data. The transformation step of the proposed algorithm must be capable of transforming any prior model realisations, not just those created using RBIG. Additionally, new observations should be taken into account during the transformation, as they may exhibit slightly different distributions.

To address this issue, the proposed algorithm simultaneously applies RBIG to both the prior neighbourhood realisations and the observations (see Eq.~\ref{eq1}). A potential drawback, however, is that it will take more time to transform a vector combining observations and blocks from all realisations. Nevertheless, previous studies of the application of RBIG in geostatistics indicated that it is much faster than other methods, such as PPMT \citep{bib14,bib17}. In this case, it took 70 seconds to transform a 9-dimensional vector combining 100 realisations of 15,561 blocks and 868 observations, giving a total of 1,556,968 rows. 

\begin{figure}[ht]%
\centering
\includegraphics[width=\textwidth]{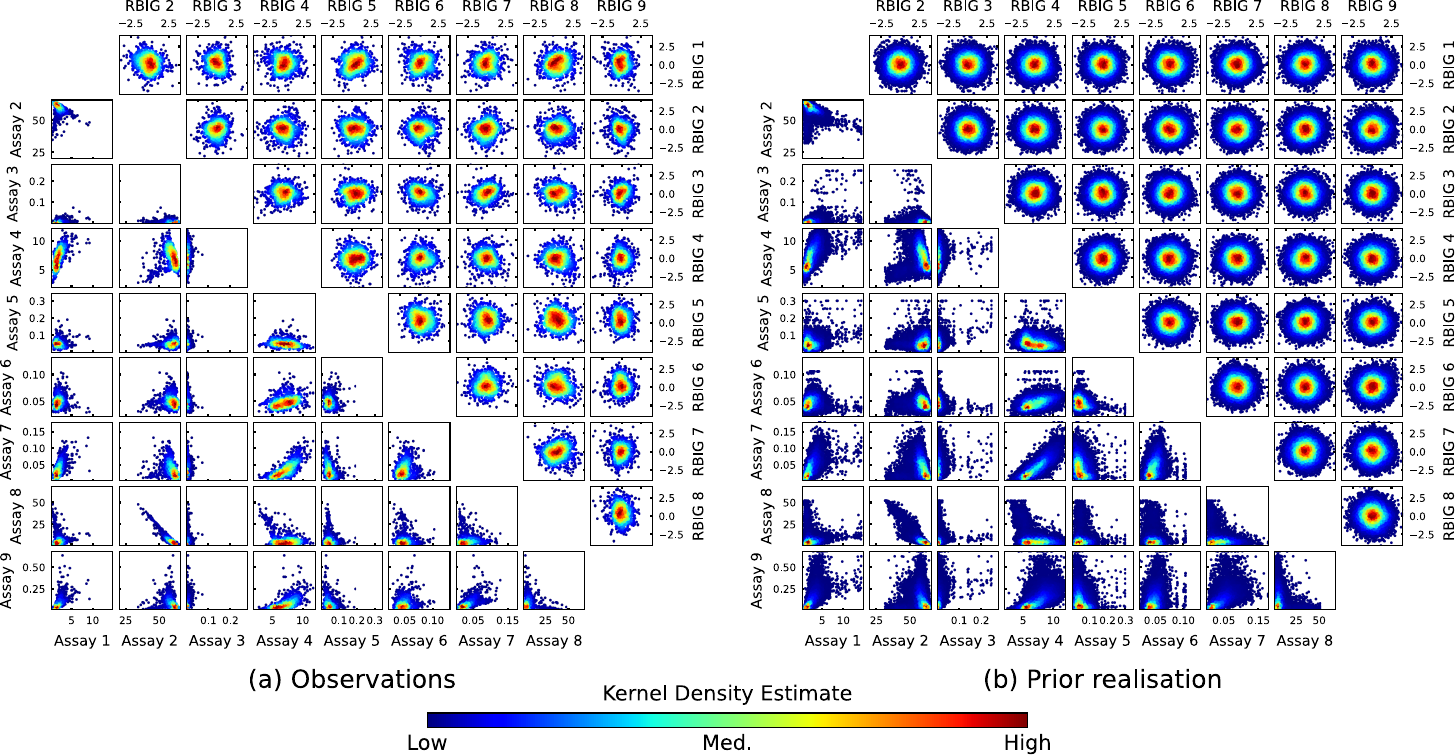}
\caption{Cross-plots of assay variables and corresponding RBIG factors for observations in period 1 and a prior realisation of the neighbourhood around the observations.}\label{fig4}
\end{figure}

Cross-plots of original and transformed variables are shown in Figure~\ref{fig4}. The multivariate relationships in this dataset are complex, with non-linearities in some bivariate distributions. There is also a noticeable skewness in some variables, particularly Assay 3. The cross plots of both prior realisation and observations are visually similar in terms of kernel density and ranges. Ultimately, RBIG successfully generated multi-Gaussian factors that satisfy the Gaussian assumption of EnKF-MDA.

Multi-Gaussian realisations and observations are used as inputs for EnKF-MDA to perform rapid updating. The critical decision at this stage is to choose the number of data assimilations, as this choice impacts both performance and computation time. Typically, the run time of EnKF-MDA is equal to the run time of EnKF multiplied by the number of data assimilations. However, it is important to note that while accuracy improves with each data assimilation, the rate of improvement decreases with each subsequent assimilation. Table~\ref{tab3} shows the mean squared error (MSE) reduction results for all variables updated with a number of data assimilations ranging from 1 to 10. MSE reduction is defined as

\begin{equation}
\text{MSE}\downarrow=\frac{\text{MSE}_\text{before}-\text{MSE}_\text{after}}{\text{MSE}_\text{before}}\times100\%,\label{eq9}
\end{equation}
where $\text{MSE}_\text{before}$ is an error between prior predictions and observations and $\text{MSE}_\text{after}$ is an error between updated predictions and observations. 

\begin{table}[h]
\begin{center}
\caption{MSE reduction (\%) for different numbers of data assimilations in period 1.}\label{tab3}%
\begin{adjustbox}{width=1\textwidth}
\begin{tabular}{lccccccccc}
\toprule
Number & Assay 1 & Assay 2 & Assay 3 & Assay 4 & Assay 5 & Assay 6 & Assay 7 & Assay 8 & Assay 9\\
\midrule
1 & 66.12 & 66.30 & 71.78 & 60.89 & 63.26 & 63.07 & 68.45 & 62.80 & 76.56\\
2 & 76.11 & 76.20 & 80.37 & 72.31 & 74.13 & 73.98 & 77.87 & 73.92 & 83.80\\
3 & 81.05 & 80.87 & 84.19 & 77.88 & 79.75 & 79.45 & 82.36 & 80.84 & 86.96\\
4 & 84.35 & 84.27 & 87.12 & 81.75 & 83.21 & 83.02 & 85.49 & 83.34 & 89.42\\
5 & 86.61 & 86.61 & 89.12 & 84.41 & 85.59 & 85.48 & 87.64 & 85.61 & 91.09\\
6 & 88.19 & 88.16 & 90.37 & 86.24 & 87.35 & 87.22 & 89.09 & 87.47 & 92.10\\
7 & 89.45 & 89.43 & 91.42 & 87.71 & 88.71 & 88.59 & 90.27 & 88.82 & 92.98\\
8 & 90.50 & 90.55 & 92.39 & 88.97 & 89.78 & 89.72 & 91.28 & 89.77 & 93.79\\
9 & 91.33 & 91.40 & 93.10 & 89.94 & 90.65 & 90.61 & 92.06 & 90.61 & 94.38\\
10 & 92.01 & 92.11 & 93.69 & 90.76 & 91.36 & 91.34 & 92.71 & 91.28 & 94.87\\
\botrule
\end{tabular}
\end{adjustbox}
\end{center}
\end{table}

A standard EnKF helps reduce MSE by 61-77\% across nine assay variables. With five data assimilations, the error reduction improves to 84-91\%, which provides a good balance between accuracy and computation time. Finally, after ten data assimilations, all nine variables achieved an MSE reduction of over 90\%. Each data assimilation, which involved 100 realisations with 15,561 blocks and 868 observations, took 1.7 seconds to complete. This means that ten data assimilations took just 153 seconds in total. Given the relatively fast computation speed, we have set the number of data assimilations in this case to ten. Additionally, the covariance localisation radius was chosen to be 30 m.

\begin{figure}[ht]%
\centering
\includegraphics[width=\textwidth]{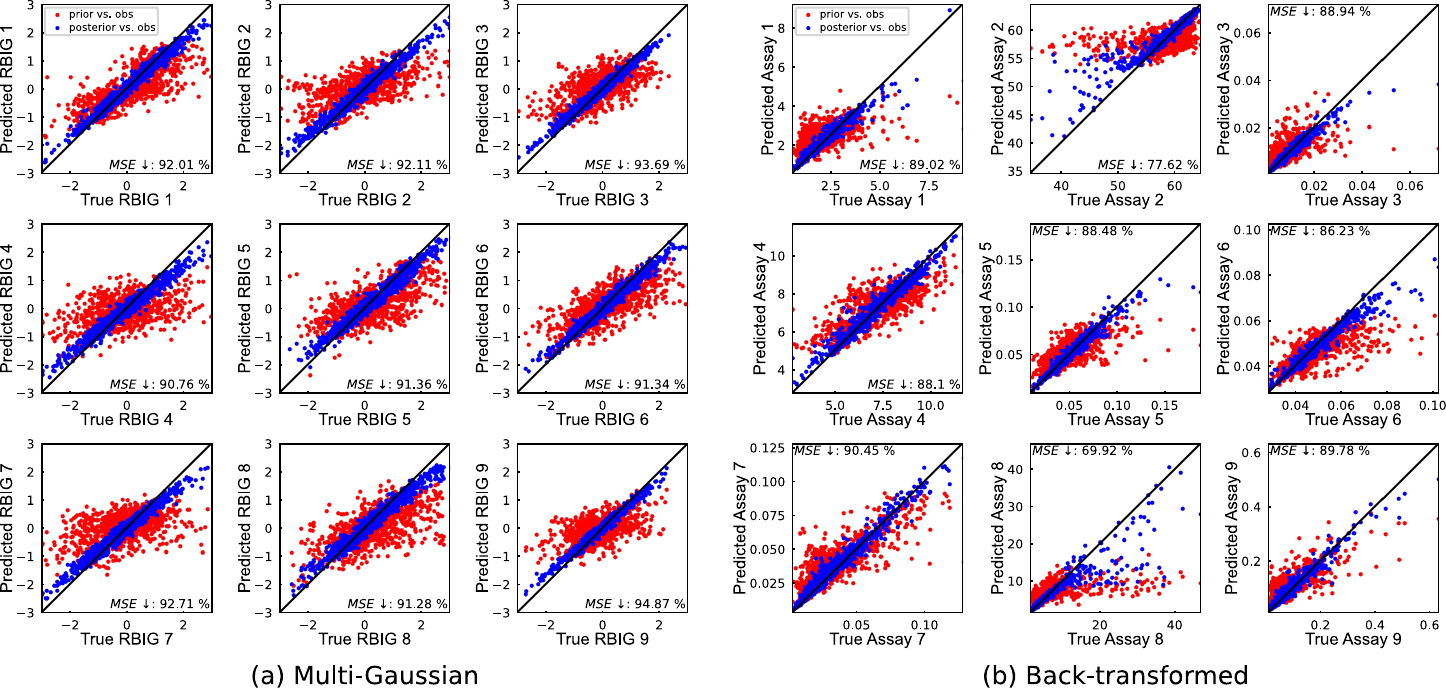}
\caption{Predictions versus observations plots before and after the update in period 1 in multi-Gaussian and back-transformed states.}\label{fig5}
\end{figure}

Predictions versus observations plots indicate that the updated multi-Gaussian results closely align with the diagonal line (Figure~\ref{fig5}). After back-transformation, all variables, except for Assay 2 and Assay 8, maintained an error reduction close to 90\%. The lower error reductions for these two variables can be attributed to the combination of different factors. As shown in Table~\ref{tab1} and Figure~\ref{fig5}, these two have larger values compared to other variables, are skewed and were predicted less accurately in the prior model. Although Assay 1 also has relatively larger values than others, its distribution is not skewed. On the other hand, despite it being significantly more skewed, Assay 3 has an excellent error reduction of 89\% due to a much closer prior prediction. The long tails of the predicted values for Assay 2 and Assay 8 contain numerous highly inaccurate predictions, ultimately impacting the final updated results. Notably, removing Assay 2 and Assay 8 doesn’t significantly affect other variables, with only a 1.24\% average difference compared to the results in Figure~\ref{fig5}.

\begin{figure}[ht]%
\centering
\includegraphics[width=0.7\textwidth]{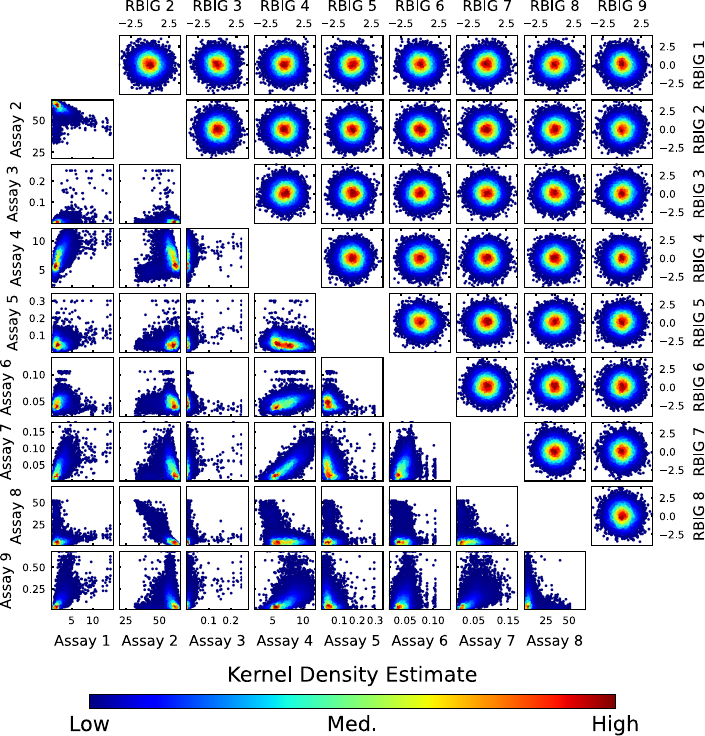}
\caption{Cross-plots of assay variables and corresponding RBIG factors for an updated realisation in period 1.}\label{fig6}
\end{figure}

The proposed approach not only accurately updated most of the variables but also preserved the multivariate relationships. In Figure~\ref{fig6}, cross-plots of RBIG factors and back-transformed assay variables for the updated realisation show distributions identical to those in Figure~\ref{fig4}. Additionally, the back-transformation of all updated neighbourhood realisations was completed in just 55 seconds. This is a major advantage of combining EnKF-MDA with RBIG, as many traditional methods struggle to maintain these complex relationships or lack computational efficiency.

All data assimilation and multi-Gaussian transformations were carried out on a Mac Mini (Apple M4 chip with four performance cores, six efficiency cores, and 16 GB of RAM). The data assimilation was implemented in Python, with covariances calculated using Cython. The multi-Gaussian transformation and back-transformation were performed in MATLAB. Overall, updating the model in period 1 took 278 seconds: 70 seconds for the RBIG transformation, 153 seconds for the EnKF-MDA, and 55 seconds for the back-transformation. This efficiency makes the proposed algorithm suitable for near real-time applications in mining, where accurate and up-to-date resource models are essential for rapid decision-making.

\subsection{Sequential rapid updating for the remaining periods}\label{subsec3.3}

To evaluate the effectiveness of the proposed rapid updating algorithm over time, the resource model realisations were updated sequentially using observations from 25 time periods. However, unlike the rapid updating in period 1, the subsequent periods have one key difference. The neighbourhood around the observations in these periods may contain blocks that were accurately updated in earlier periods. One way to mitigate the issue of potentially overwriting the previously updated blocks is to incorporate observations from prior periods. Therefore, after selecting the neighbourhood around new observations, the proposed algorithm also includes previous observations that are located in the same neighbourhood. Table~\ref{tab4} shows the MSE reduction results after updating the first five periods, comparing the scenarios with and without the inclusion of previous observations. Across all nine variables, the error reductions show consistent improvement. Additionally, since updates are performed within local neighbourhoods around new observations, including previous observations does not significantly slow down the process.

\begin{table}[h]
\begin{center}
\caption{MSE reduction (\%) after period 5 with and without incorporating previous observations.}\label{tab4}%
\begin{adjustbox}{width=1\textwidth}
\begin{tabular}{lccccccccc}
\toprule
Scenario & Assay 1 & Assay 2 & Assay 3 & Assay 4 & Assay 5 & Assay 6 & Assay 7 & Assay 8 & Assay 9\\
\midrule
With & 89.46 & 74.02 & 84.77 & 88.38 & 89.14 & 89.29 & 90.85 & 65.49 & 91.26\\
Without & 91.95 & 78.65 & 93.53 & 90.81 & 92.24 & 92.22 & 92.89 & 70.97 & 92.85\\
\botrule
\end{tabular}
\end{adjustbox}
\end{center}
\end{table}

The following figures illustrate how the model is progressively refined as more observations become available. Instead of presenting the updates for each individual period, the updates are grouped into approximately equal batches of observations, focusing on a 2D section of the model at an elevation of 44 m.

\begin{figure}[ht]%
\centering
\includegraphics[width=0.65\textwidth]{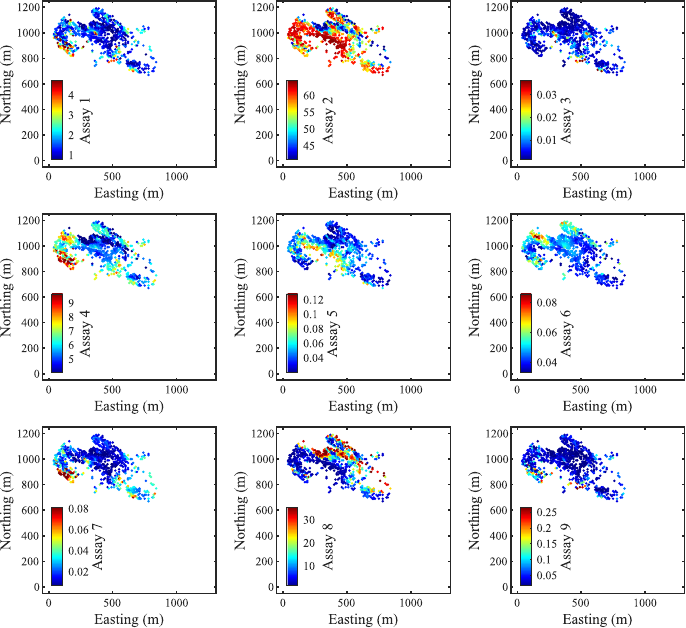}
\caption{2D view of observations from periods 1-5 at 44 m elevation.}\label{fig7}
\end{figure}

The prior models, shown in Figure~\ref{fig3}, exhibit a smoothing effect that is common in geostatistical modelling. Figure~\ref{fig7} shows the observations from periods 1 to 5 at an elevation of 44 m. In contrast to prior maps, these observations are evidently different at the exact locations and exhibit greater spatial variability. This emphasises why rapid updating is important, as resource models often struggle to capture small-scale variability due to the limited resolution of exploration data. Although sensor observations come with a degree of uncertainty, they offer a vast amount of real-time data that can be integrated into the model.

\begin{figure}[ht]%
\centering
\includegraphics[width=0.65\textwidth]{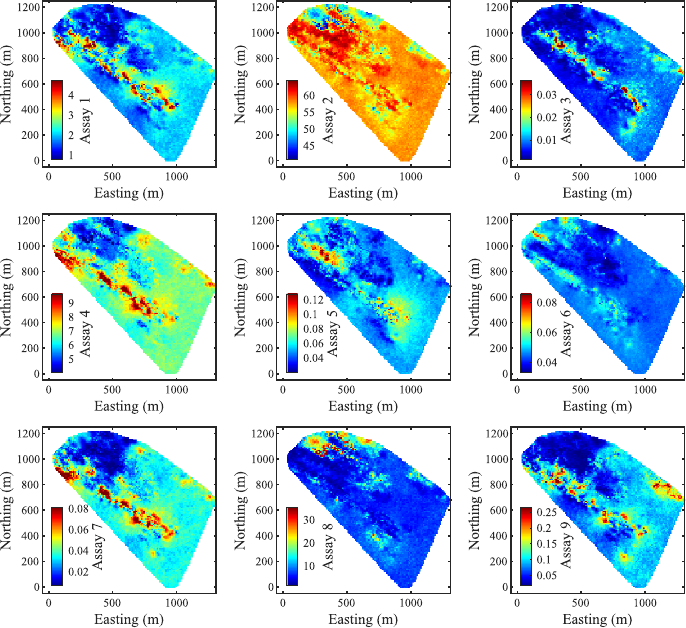}
\caption{2D view of updated resource models after period 5 at 44 m elevation.}\label{fig8}
\end{figure}

After updating the model sequentially over the first five periods, the updated e-type models are shown in Figure~\ref{fig8}. Firstly, the updated models align more closely with the observations and demonstrate greater variability in that part of the deposit. The rapid updating also increased the spatial variability and reduced the over-smoothing in the area surrounding the observations. However, there is a region between 600-800 m northing and 1100-1300 m easting that differs significantly from the prior models despite lacking observations there. This is likely the result of neighbouring observations at adjacent elevations affecting this area.

\begin{figure}[ht]%
\centering
\includegraphics[width=0.65\textwidth]{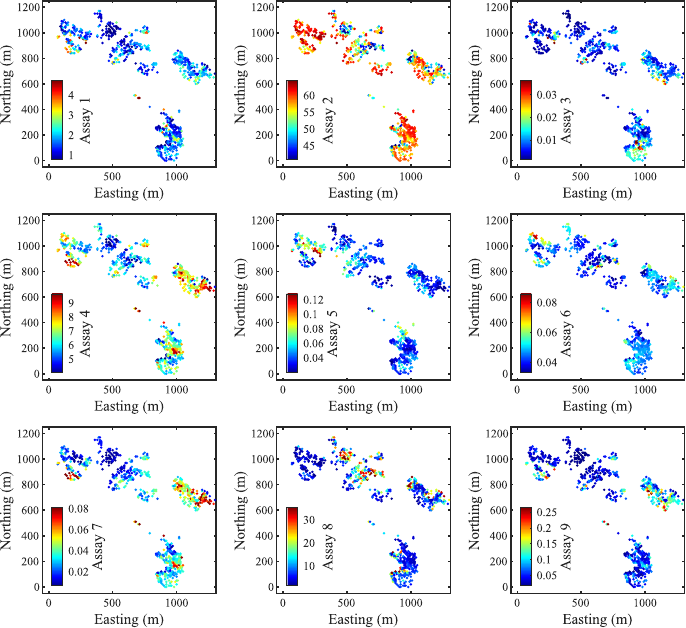}
\caption{2D view of observations from periods 6-9 at 44 m elevation.}\label{fig9}
\end{figure}

\begin{figure}[ht]%
\centering
\includegraphics[width=0.65\textwidth]{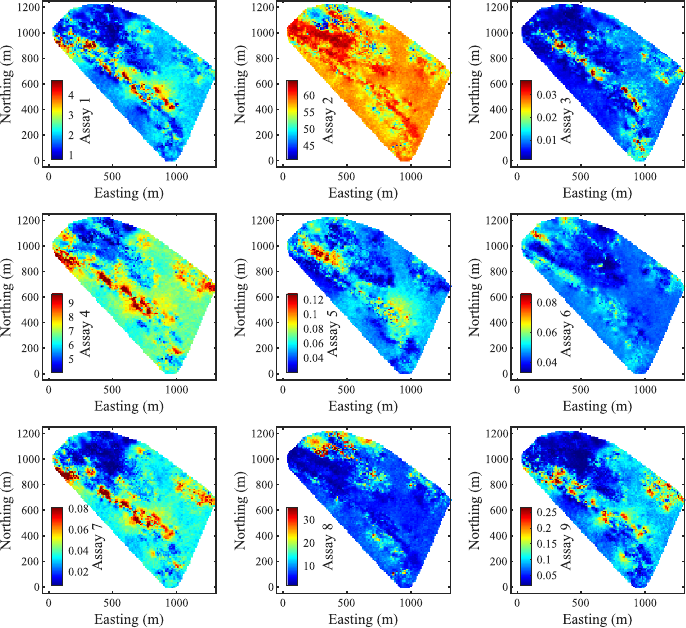}
\caption{2D view of updated resource models after period 9 at 44 m elevation.}\label{fig10}
\end{figure}

Figure~\ref{fig9} illustrates four more sets of observations from periods 6 to 9. The first thing to notice here is that the area between 600-800 m northing and 1100-1300 m easting, which was unexpectedly updated in Figure~\ref{fig8}, aligns closely with actual observations at those locations. This indicates that nearby data can contribute to refining the model predictions. However, the biggest discrepancies between new observations and previous models are found in the lower sections of the maps. Updated models after period 9 now provide a more accurate representation of that lower part of the map (Figure~\ref{fig10}). Furthermore, previously underestimated high-grade zones have also become more distinct for variables such as Assay 4 and Assay 7.

\begin{figure}[ht]%
\centering
\includegraphics[width=0.65\textwidth]{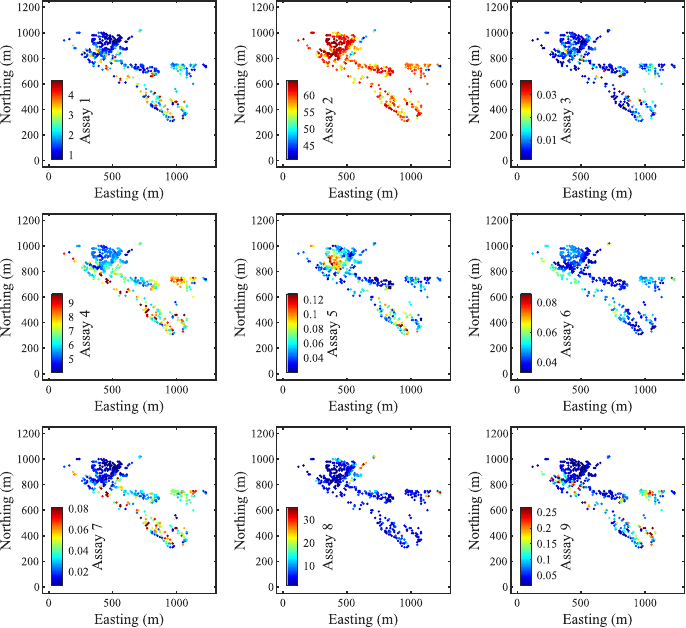}
\caption{2D view of observations from periods 10-25 at 44 m elevation.}\label{fig11}
\end{figure}

\begin{figure}[ht]%
\centering
\includegraphics[width=0.65\textwidth]{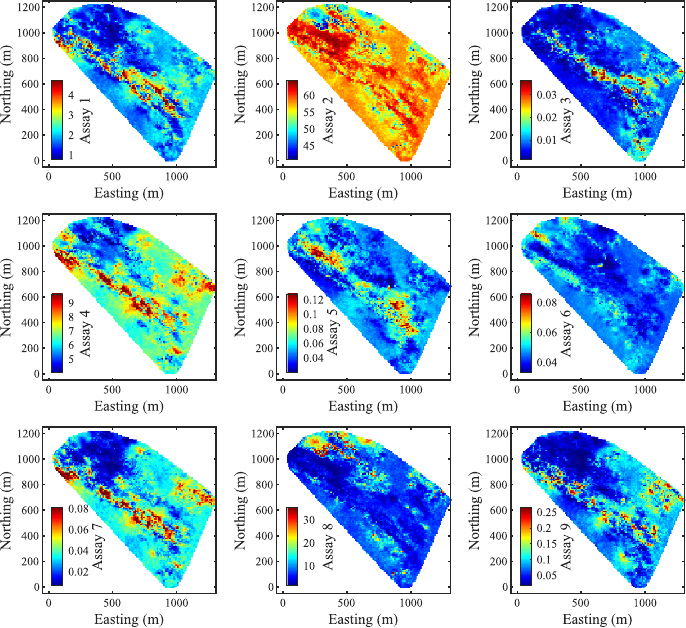}
\caption{2D view of final updated resource models at 44 m elevation.}\label{fig12}
\end{figure}

Finally, the observations from the remaining periods are presented in Figure~\ref{fig11}. The reason for displaying such a large set of periods is that most observations from this batch are not present at an elevation of 44 m. The final updated models demonstrate greater spatial variability and a higher level of detail compared to the prior models, particularly in areas with dense observation coverage (Figure~\ref{fig12}). Visually, there is a clear and gradual improvement throughout the updates, as seen in Figures~\ref{fig8} and \ref{fig10}.

\begin{figure}[H]%
\centering
\includegraphics[width=0.7\textwidth]{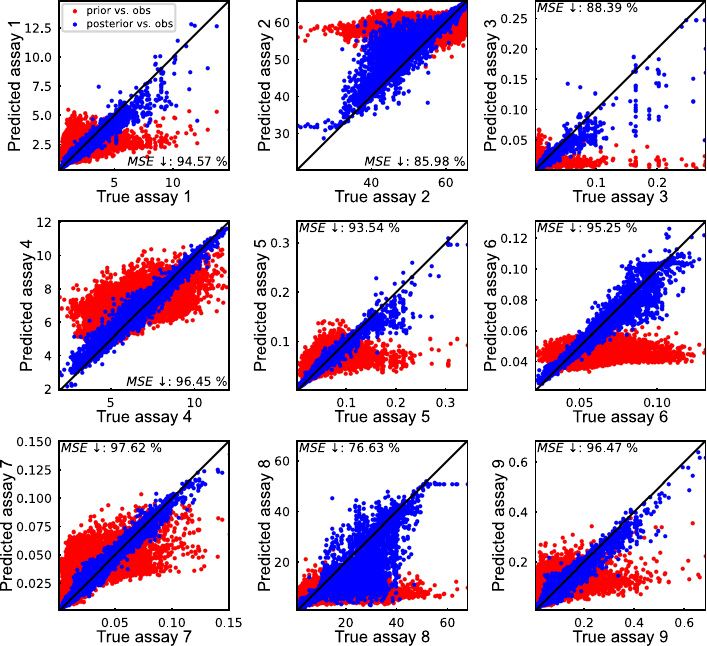}
\caption{Predictions vs. observations plots for prior resource models (in red) and after the final update in period 25 (in blue).}\label{fig13}
\end{figure}

A more detailed analysis of the final updated models is presented in Figure~\ref{fig13}, where predictions are plotted against the observed values. The updated results closely align with the diagonal line, and most variables achieve an error reduction between 86\% and 9\%. However, Assay 8 only reduced its MSE by 76.73\%, which is significantly lower compared to the other variables. This discrepancy is particularly evident in the plot, with many points positioned far from the diagonal line.

A similar trend was noted in Figure~\ref{fig5}, where, for period 1, Assay 8 achieved an error reduction of less than 70\%. As mentioned earlier, this issue can be partly attributed to the skewness of the distribution, which makes it challenging to accurately back-transform the tail end. Other factors are the relatively high number of significantly inaccurate prior predictions and the larger value range.

Interestingly, Assay 2 showed an improvement in error reduction compared to period 1, but there are still points that deviate from the diagonal line. Assay 3 demonstrates even better accuracy despite having a more skewed distribution. The difference between Assay 8 and Assay 3 is that the latter has fewer observations at the tail end of its distribution. However, RBIG still struggled to back-transform the tails in both cases, raising concerns about its reliability for highly skewed distributions. Overall, the proposed algorithm effectively reduced the MSE by 94-98\% across six out of nine variables.

\begin{figure}[ht]%
\centering
\includegraphics[width=\textwidth]{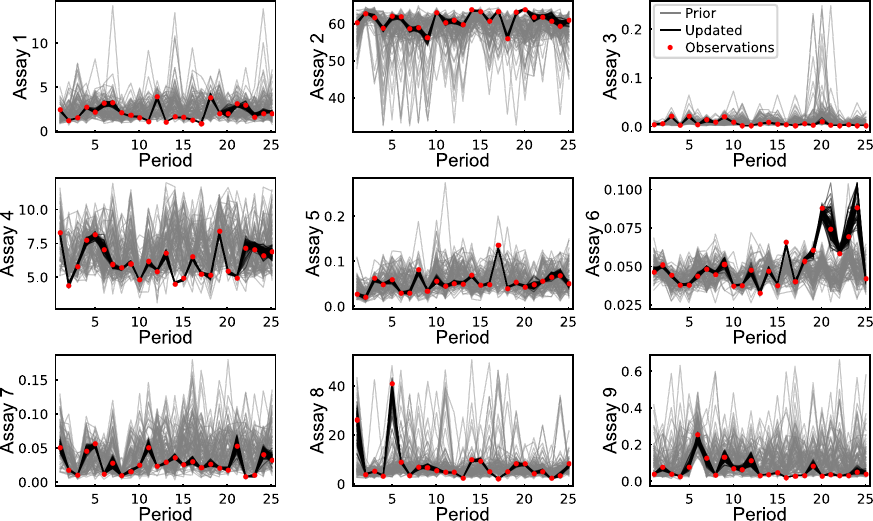}
\caption{Visualisation of realisations before and after rapid updating for blocks with errors close to the median in each period.}\label{fig14}
\end{figure}

All the previous figures primarily analyse the average results from all realisations. To focus on how rapid updating affects individual realisations, we calculated the errors between the updated results and the observations. In each period, we selected a block with an error close to the median of all the errors. We chose the median because it ensures that 50\% of the blocks exhibit better predictions while the other 50\% show worse predictions. Figure~\ref{fig14} displays the prior and updated realisations for the blocks that have errors near the median in each period. The updated results closely match the observations for each variable. Furthermore, rapid updating has minimised uncertainty, as evidenced by the reduced spread of realisations before and after the updates. However, uncertainty is slightly higher for larger values due to the skewness of the distributions. This is particularly evident in Assay 6, where values in later periods are significantly higher and have a broader spread of realisations. Interestingly, Assay 8, despite showing relatively low error reduction, shows very accurate results in this figure.

\begin{figure}[ht]%
\centering
\includegraphics[width=\textwidth]{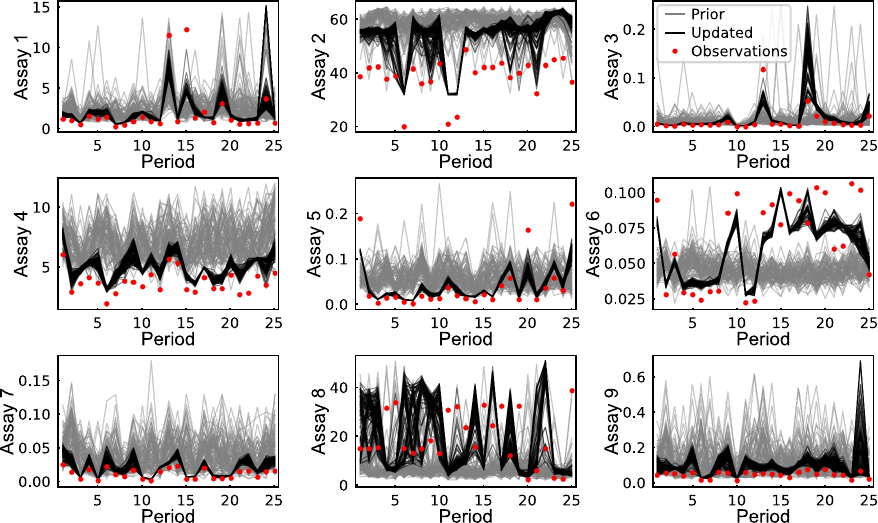}
\caption{Visualisation of realisations before and after rapid updating for blocks with highest errors in each period.}\label{fig15}
\end{figure}

The proposed approach, as shown in the predictions versus observations plots in Figure~\ref{fig13}, produces some outliers that are far from the diagonal line. This issue is particularly evident in Assays 2 and 8, which demonstrate lower error reductions than the other variables. To illustrate this, Figure~\ref{fig15} shows the realisations before and after rapid updating for the blocks with the highest errors during each period. Variables that had good error reductions remain close to the observed values. For instance, in Assay 4, the slight deviation from the observations is primarily due to overestimation in prior models. In contrast, Assay 6 shows that prior models have both overestimation and underestimation in different periods, leading to deviations from observations in the updated models. For Assays 2 and 8, however, the gap between realisations and observations is more significant, primarily due to back-transformation issues.

\section{Discussion}\label{sec4}

The application of the proposed combination of EnKF-MDA and RBIG in a real case study demonstrated its effectiveness in rapidly and accurately updating multivariate resource models. Across all nine variables, the proposed approach achieved an error reduction ranging from 77\% to 98\%, reducing errors by more than 94\% for five of those variables. The results indicate that the updated models provide a more realistic representation of spatial variability and align closely with observations while also maintaining multivariate relationships. Moreover, transforming realisations and observations together helps to account for potential new observations coming from previously under-sampled locations.

In addition to the accuracy, the proposed algorithm operates with impressive speed, even on low-cost hardware. For example, it can update 100 block model realisations with 878 observations in under five minutes. The process could be further enhanced through parallelisation across multiple virtual machines, which is easily achievable in an industrial setting. Notably, the rapid updating requires no human intervention and can be automated to run either at specified time intervals or immediately when new information becomes available. As a result, mining operations can make near real-time decisions based on updated models, leading to improved short-term mine planning and optimisation.

Achieving such computational efficiency and flexibility comes with some drawbacks and limitations. Despite its advantages the Kalman filter (KF) is less effective than kriging or cokriging estimates. Li et al. \citep{bib23} compared the updates from KF to estimates produced by ordinary kriging, which unsurprisingly showed better performance for kriging. Incorporating kriging to update resource model realisations can be done by using residuals between observations and prior realisations \citep{bib21,bib34}. However, such kriging updates are still computationally demanding and inflexible to be used in the real-time mining framework presented in Figure~\ref{fig1}. On the other hand, data assimilation methods are faster, especially when dealing with large resource models. The use of EnKF-MDA helps to further minimise the deviation between model predictions and observations by updating the same data multiple times.

Another drawback noted in this paper is the less accurate updating of extreme values in highly skewed distributions. This limitation is characteristic of RBIG and similar transforms, which can produce artefacts in the presence of extreme values. On the contrary, FA has previously been used for rapid updating \citep{bib10} and can minimise artefacts after back transformation. However, it is also significantly more computationally intensive and using it for time-sensitive tasks such as rapid updating is not recommended \citep{bib14,bib17}. Additionally, multi-Gaussian transforms require the multivariate dataset to be homotopic, making the proposed approach impractical for datasets where some variables are under-sampled. In such cases, a data imputation technique such as Gibbs sampling can be used to generate additional observations \citep{bib35,bib36}.

Finally, the real data used in the case study has already undergone data fusion and ore tracking. For confidentiality reasons, measurement errors and ore tracking uncertainty were not disclosed. For simplicity, a measurement error of 10\% is assumed in this paper. However, the uncertainty associated with sensor observations extends beyond measurement errors typically identified in laboratory settings. The way sensors are used on an actual mine site can differ from how they are intended to be used. Not to mention the difference between harsh mining conditions and a laboratory environment in which measurement errors are evaluated. Furthermore, although more practical, downstream sensors, such as those on conveyor belts, are hard to link back to resource models. This challenge is particularly relevant in underground mining, where ore tracking remains a significant issue. Thus, the uncertainty introduced by ore tracking must be considered during real rapid updating.

\section{Conclusions}\label{sec5}

This paper presents a rapid updating algorithm designed explicitly for multivariate resource models. The algorithm combines EnKF-MDA and RBIG, where the former offers more accurate updates compared to the traditional EnKF, while RBIG transforms multivariate data into independent multi-Gaussian factors that are suitable for updating. This approach is applicable to any mining deposit that has multiple cross-correlated quantitative variables. The effectiveness of the proposed algorithm was demonstrated through a real case study of an iron ore mine in Western Australia.

Quantitative grade variables are not the only critical components of resource models. Qualitative (e.g., lithology, alterations and various domain types) and non-additive geometallurgical variables (e.g., operating work index, ore recovery, Axb, etc.) are also essential for resource modelling, optimisation, and mine planning. The primary challenge with these types of variables is that ensemble-based data assimilation mainly operates on Gaussian values. However, EnKF has been applied to update geological domains by using discrete wavelet transforms \citep{bib37}. Additionally, EnKF has been effective in updating the Bond Ball Mill Work index and improving its future forecasts by 26\% \citep{bib9}.

Future research will focus on expanding the rapid updating algorithm to enable the updating of qualitative variables and geometallurgical properties. The limitations of the current approach will be investigated further, especially the updating of extreme values in highly skewed distributions. Finally, it is important to test how data assimilation performs in a real underground mining scenario and its applicability in enhancing short-term mine planning.

\backmatter

\bmhead{Acknowledgments} 

The research reported here was supported by the Australian Research Council Industrial Transformation Training Centre for Integrated Operations for Complex Resources (ARC ITTC IOCR - project number IC190100017) and funded by universities, industry and the Australian Government. We acknowledge Petra Data Science for providing a fused dataset for the case study.

\bibliography{bibliography}% common bib file

\end{document}